\documentclass[graybox]{svmult}

\usepackage{mathptmx}       
\usepackage{helvet}         
\usepackage{courier}        
\usepackage{type1cm}        
\usepackage{makeidx}         
\usepackage{graphicx}        
\usepackage{multicol}        
\usepackage[bottom]{footmisc}

\makeindex             

\begin{document}

\title{The Nordic Optical Telescope}
\author{Anlaug Amanda Djupvik and Johannes Andersen}
\institute{Anlaug Amanda Djupvik \& Johannes Andersen 
\at Nordic Optical Telescope, Apdo 474, 38700 Santa Cruz de La 
Palma, Spain, \email{amanda@not.iac.es, ja@not.iac.es}
}
%
%
\maketitle

\abstract{An overview of the Nordic Optical Telescope (NOT) is presented. 
Emphasis is on current capabilities of direct interest to the scientific 
user community, including instruments. Educational services and prospects 
and strategies for the future are discussed briefly as well.}

\section{Introduction}
\label{sec:intro}

The Nordic Optical Telescope (NOT) is a modern 2.6m alt-azimuth telescope, 
operating at Roque de los Muchachos Observatory, La Palma, since 1989 as 
the main northern-hemisphere optical facility for Nordic astronomers. Its
f/11 Ritchey-Chretien optical system, the telescope, and the enclosure 
were carefully designed to deliver the best image quality at the site 
(Ardeberg, 1990).

{\em Structure.}
The Nordic Optical Telescope Scientific Association (NOTSA) was formed 
in 1984 to construct and operate the NOT on behalf of the Research Councils 
of Denmark, Finland, Norway, and Sweden; the University of Iceland joined 
in 1997. The budget is shared roughly in the ratios 20:30:20:30:1\%. 

A Director has overall legal, financial, and scientific 
responsibility for operations and strategic planning, under the authority
of the NOTSA Council, which represents the owners. On-site operations are 
ensured by a team of 13 staff members headed by an Astronomer-in-Charge; 
5 research students complement the team. 

\begin{figure}[t]
\center{
\includegraphics[scale=0.4]{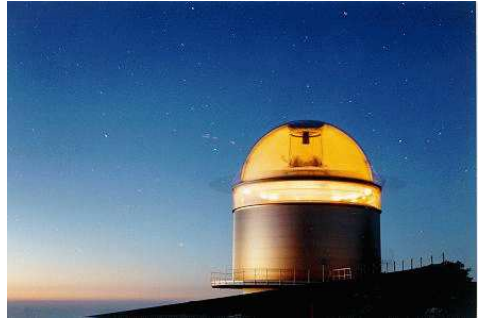}
\caption{The NOT at sunset with the dome lights on while turning 
(Photo: Jacob Clasen, NOT).}
}
\label{fig:bluesky}
\end{figure}

{\em User Community.}
\label{sec:users}
75\% of the science time on the telescope are at the disposal of the 
Nordic community through competitive proposals peer-reviewed by the 
NOT Observing Programmes Committee. 20\% of the time are allocated to 
Spanish astronomers through the Comisi\'on de Asignaci\'on de Tiempo 
(CAT), while 5\% go to international teams through the CCI International 
Time Progamme (ITP). 

Not only Nordic astronomers, but {\em anybody} regardless of nationality 
or affiliation can apply for the Nordic time, and proposals are ranked 
solely on the basis of scientific merit. Thus, over the past 5 years, over 
20\% of the Nordic time has been allocated by the OPC to non-Nordic PIs. 
About half of these projects have been part of the OPTICON Trans-National 
Access Programme, which supports access for European researchers to all 
modern European 2-4m class telescopes. Thus, despite its name, the NOT 
already serves a broad European user community.


\section{NOT User Services}
\label{sec:not}

A systematic upgrade programme has been conducted over the last five years
to improve the scheduling, operation, instrumentation, and reliability of 
the NOT. 

{\em Scheduling.}  
The NOT offers both visitor and service mode observations. The emphasis is 
increasingly on the latter because of the flexibility offered, and the 
fraction  of service observing nights has increased steadily to $\sim$40\% 
in 2005 -- 2007. 

Part of the service time is allocated through a ``Fast-track'' procedure 
with a short lead time from proposal to execution, offered since 2005. 
Proposals for short observing programmes (max. 4 hours) can be can submitted 
at any time (see http://www.not.iac.es/observing/service/) and are reviewed 
promptly by the OPC. The successful programs are then executed by the staff 
in queue scheduled service mode. Available instruments are ALFOSC, NOTCam, 
FIES and StanCam. 

The pressure factor from proposals for the Nordic time has increased gradually 
in recent years, to about 2.5 over the last decade. Technical downtime is 
typically $\sim$1\%, weather down time $\sim$10\% in summer (April-September) 
and $\sim$35\% in winter (October-March). However, actual usable hours vary 
by only $\sim$10\% over the year.

{\em Operations.}  
Major effort has been put into making the telescope control system fast, 
flexible, safe and easy to use. Telescope operators are not needed, but 
there is always a support astronomer to provide a thorough introduction 
on the first night of a visitor run, and technical support is at hand 
throughout. 

For safety, the telescope is linked to our weather
station, and the dome closes automatically if limits of humidity 
or wind are exceeded. Autoguiding starts fully automatically in 
most observing modes, and focusing is fast and reliable. 

The entire observing system (detector + instrument + telescope) can be 
fully controlled by the new integrated data acquisition system, the 
{\em Sequencer}, implemented for the core instruments in 2006-2008. 
This permits in principle to automate the entire observation, 
although these features are currently under development.

Only one instrument at a time can be mounted at the main focus, but
45-degree folding mirrors in the adaptor allow the standby CCD camera 
(StanCam) and the high-resolution fiber spectrograph (FIES) to be 
available at all times, greatly increasing flexibility on time-critical 
programmes.

A rare feature of the NOT is its ability to go as low as 6.4$^{\circ}$ 
from the horizon, which is a unique capability in studies of objects in 
the inner Solar System, such as the planet Mercury to (see e.g. 
Warell \& Karlsson, 2007).

{\em Documentation and Data flow.}  
A comprehensive set of documentation on the telescope, instruments, 
detectors, and operations is available at http//www.not.iac.es. This  
includes on-line tools for preparing proposals and/or observations, 
such as an Exposure Time Calculator and Script Generators for the core 
instruments. Quality control data on detectors etc. are also available. 

All data from the core instruments (ALFOSC, FIES, MOSCA, NOTCam and
StanCam) are stored in standard Multi-Extension Fits (MEF) format with 
primary WCS information in the headers. The data are delivered on DVD, 
or available by ftp for ``Fast-Track'' or time-critical Target-of-Opportunity 
observations. Technical details of the core instruments are summarised in 
the following sections. The optical zero-point magnitudes are summarized in
Table~\ref{tab:1}.

\begin{table}
\caption{UBVRI zero-point magnitudes (for 1e$^-$/s).}
\label{tab:1}
\begin{tabular}{p{3cm}p{1.5cm}p{1.5cm}p{1.5cm}p{1.5cm}p{1.5cm}}
\hline\noalign{\smallskip}
           & $U$ & $B$ & $V$ & $R$ & $I$ \\
\noalign{\smallskip}\svhline\noalign{\smallskip}
ALFOSC     & 24.0 & 25.7 & 25.6 & 25.4 & 24.6 \\
MOSCA      & 24.6 & 26.2 & 26.0 & 25.7 & 25.2 \\
StanCam    & 23.8 & 25.7 & 25.4 & 25.3 & 24.5 \\
\noalign{\smallskip}\hline\noalign{\smallskip}
\end{tabular}
\end{table}

\section{ALFOSC}

The Andalucia Faint Object Spectrograph and Camera (ALFOSC), owned by 
the Instituto de Astrof\'isica de Andaluc\'ia (IAA), has been used at 
the NOT by mutual agreement since 1996. ALFOSC was upgraded with a new 
CCD detector (e2v CCD42-40, 13.5$\mu$m $\times 2048 \times 2048$) 
and new camera optics in 2003. 
The plate scale is 0.19''/pix. The new CCD provides excellent resolution 
at all wavelengths, but higher fringe levels than the previous CCD. 
Fast photometry with windowed readout and on-line reduction with 
comparison star subtraction is offered.

With 16 grisms, ALFOSC offers spectroscopy in the resolution range $200 < 
R < 10000$. A VPH grism with spectral resolution R=10000 in the 6350-6850 
\AA \ range offers a total system efficiency of 30\%.
A set of vertical slits allow fast horizontal readout in a 
small window for time-resolved spectroscopy, and quick-look reduction tools 
are available for both long-slit and echelle spectra.
Multi Object Spectroscopy (MOS) with pre-fabricated slit plates is available.

Dual-beam linear and circular imaging- and spectro-polarimetry is also
offered with ALFOSC. A quick-look reduction tool for linear imaging 
polarimetry was added in 2007 
(see http://www.not.iac.es/instruments/alfosc/).

\begin{figure}[t]
\center
\includegraphics[scale=0.45]{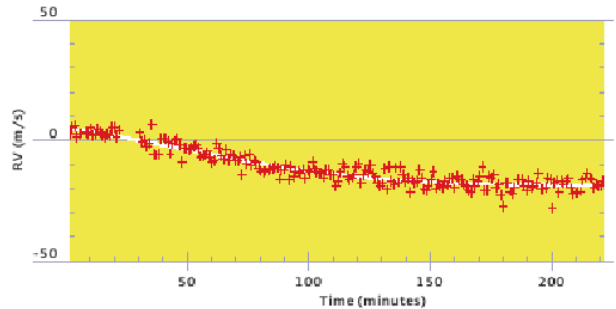}
\caption{A series of daytime sky spectra (reflected sunlight). The mean 
trend is likely due to winds in the upper atmosphere; the dispersion 
around the mean is $<$~3~m/s.}
\label{fig:bluesky}
\end{figure}

\section{FIES}

The FIber Echelle Spectrograph (FIES) is permanently ready for use. For 
mechanical and thermal stability it is located in a separate building 
with thermal control. The detector is an e2v CCD42-40 
(13.5$\mu$m $\times 2048 \times 2048$), which covers the wavelength 
range 370-730 nm without gaps. The fiber diameters are 2.5'' in 
low-resolution mode (R = 25000), 1.3'' for medium (R = 45000) and 
high resolution mode (R = 65000; slit added). The FIEStool package 
provides on-line quick-look spectra and is also suitable for final 
reductions.

A main goal for FIES is high-precision radial-velocity work. The 
zero-point stability is currently $<$~150~m/s for a stellar spectrum 
followed by a separate Th-Ar lamp exposure, $<$~15~m/s for a stellar 
spectrum with {\em simultaneous} Th-Ar calibration. As an example, the 
3 M$_{\rm Jup}$ transiting planet WASP-10b was confirmed i.a with 
radial velocities from FIES (Christian et al. 2008). Tests with series 
of daytime blue-sky spectra using simultaneous Th-Ar are also giving 
promising results (see Fig.~\ref{fig:bluesky}; Frandsen et al. 2007). 
An iodine absorption cell is being added.

FIES is rapidly attracting interest and became the most-demanded 
instrument at the NOT among the applications for the semester 2008B.
For more details, see http://www.not.iac.es/instruments/fies/.

\section{NOTCam}

The Nordic Optical Telescope near-infrared Camera and spectrograph 
(NOTCam) is a versatile instrument for the 0.8 - 2.5 $\mu$m 
wavelength range, using a recent Rockwell Hawaii-I HgCdTe detector 
(18$\mu$m $\times 1024 \times 1024$). 

\begin{figure}[t]
\sidecaption
\includegraphics[scale=0.32]{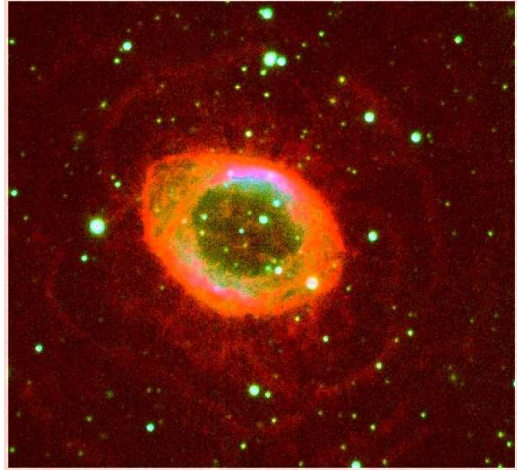}
\caption{NOTCam image of M57 in $J$ (blue), $H$ (green), and a
narrow-band filter centred on the $H_2$ line at 2.121$\mu$m 
(red). FOV is 3.5' $\times$ 3.2', north up, east left. 
Beam-switch observations with a total on-source integration time 
of 360 seconds per filter. Reduced with the NOTCam quick-look 
reduction package.}
\label{fig:m57}
\end{figure}

The image scale can be toggled in a matter of seconds 
between the wide-field (WF) camera (0.234''/pix, FOV = 4') and the 
high-resolution (HR) camera (0.078''/pix, FOV = 80''). The 
WF camera suffers from some optical distortion in the corners, 
but the HR camera has a high optical quality all over the FOV. With 
proper telescope tracking, the HR camera regularly delivers perfectly 
round, deep stellar images with FWHM of 0.3-0.4''.

Standard broad band filters $JHK$ plus $Y_n$ and 17 narrow-band 
filters are permanently installed. A cold shutter permits short 
integrations, and small cold stops are available for flux reduction 
of very bright targets.
The zero-point magnitudes (for 1 e$^-$/s) are: 24.1, 24.1, and 23.5 
mag, in $J$, $H$ and $K_S$, respectively (Vega magnitudes). An example 
image is shown in Fig.~\ref{fig:m57}. 

Spectroscopy with the WF camera and a 0.6'' slit gives a resolution 
of R = 2500 and covers the Yn, J, H and K bands.
The HR camera yields R = 5500 with a 0.2'' slit and covers wavelength 
ranges from 1.26-1.34 $\mu$m (Pa~$\beta$), 1.57-1.67 $\mu$m,
to 2.07-2.20 $\mu$m (Br~$\gamma$). 
Wollaston prisms for polarimetry, low resolution grisms (Telting, 2004), 
and broad-band ZY-filters are being considered for the future.
For more detail see http://www.not.iac.es/instruments/notcam/.

\section{MOSCA}

The MOSaic CAmera (MOSCA) is a direct imager featuring 4 Loral CCDs 
of 15$\mu$m $\times 2048 \times 2048$ corresponding to 0.11''/pix and 
a FOV of 7.7'. The gaps between the four CCDs are 9-12''. 
The advantages of MOSCA are the uniformly good PSF over the entire 
FOV and the high throughput (see Tab.~\ref{tab:1}), especially in the 
$U$ band. For more detail see http://www.not.iac.es/instruments/mosca/.

\section{StanCam}

The Stand-by Camera (StanCam) has a TEK 24 thinned CCD 
($\mu$m $\times 1024 \times 1024$; 0.176''/pix, FOV = 3'). It is 
permanently available at a folded Cassegrain focus. Together with 
NOTCam it offers near-simultaneous UBVRIJHK coverage. 
For more detail see http://www.not.iac.es/instruments/stancam/.

\section{Visitor instruments}

The Turku photo-polarimeter Turpol provides linear and circular 
simultaneous UBVRI polarimetry of single sources (diaphragms) with high 
precision through a double image chopping technique. Polarization levels 
below 0.01\% can be detected at the NOT, and systematic errors for 
brighter stars are $\sim$ 0.005\% (Piirola, 1999). On-line reduction is 
available. See http://www.not.iac.es/instruments/turpol/.

SOFIN, a high resolution echelle spectrograph, is supported by the 
University of Helsinki and Astrophysikalisches Institut Potsdam. It 
covers the spectral range
3500 - 9000 \AA \ with resolutions R = 27,000 - 170,000, and is also 
capable of spectro-polarimetry (Ilyin, 2000). Observations are 
performed in service mode and reduction
software is available. See http://www.not.iac.es/instruments/sofin/.

The PolCor ``Lucky'' polarimeter/coronagraph, equipped with an EMCCD 
(16$\mu$m  $\times 512 \times 512$, 0.12''/pix, FOV = 1') with up to
33 Hz readout rate, has a rapidly turning polarizer and three
coronagraphic disks. A computer controlled Lyot stop masks the M2 
support vanes. 
The fwhm improves from initially 0.7'' to 0.4'' applying only shift 
and add techniques, but improves further to  0.2-0.3'' using frame 
selection and deconvolution 
(Olofsson \& Flor\'en, 2008). 

Among other visitor instruments, {\em LuckyCam} was used at the NOT in 
2002-2004 to obtain near diffraction-limited imaging in the $I$ band 
using reference stars as faint as 16.5 mag. Keeping the 1\% best frames 
improved the image quality from 0.7'' to 0.1'' (Mackay et al. 2003). 
Recently, also FastCam (Oscoz \& Rebolo et al. 2007) also obtained
diffraction-limited $I$ band imaging using 1-5\% of the frames.


\section{The educational role of the NOT}
\label{sec:edu}

In a world of 8-m telescopes it makes sense to spend a modest amount of 
time on a 2m-class facility to train the next generation of astronomers –-
at least those who will be involved in designing and building the next 
generation of telescopes and their instrumentation. At the NOT, we are 
taking a systematic approach to this challenge and are developing a varied 
set of offers of educational services to universities, covering the whole 
educational ``food chain''. 

The traditional use of the NOT in education has been for on-site courses, 
where groups of typically 12 PhD or MSc students spend 1-2 weeks on La Palma 
learning how to observe with the NOT, reduce and analyse their data, and 
formulate their next observing time applications effectively. Such courses 
are highly educational, motivating, and popular with the students, giving 
them real hands-on observing experience. The NOT supports these courses by 
providing a small ``class room'' at ORM, with the control room screens 
projected on the wall and with 12 standard laptops available for data 
reduction and analysis.

These courses have become very popular, but cannot handle large 
volumes of students. We are therefore also developing remote observing as 
a means to involve hands-on observing in university courses without actually 
taking the students to La Palma. The observing courses at Mol\.{e}tai 
Observatory in Lithuania have pioneered this technique, most recently in 
2008. 

Finally, the NOT currently hosts 5 PhD or MSc students, who typically spend 
a year at on La Palma, devoting 75\% of their time to their thesis projects 
and 25\% on support duties, service observing, or various developments. 
Experience shows that this experience is a significant asset for their future
careers, whether in or outside astronomy. 


\section{Future perspectives}
\label{sec:fut}

The world of today and tomorrow is scientifically, technically, financially,
and politically very different from that of 1984 when NOTSA was founded. The 
great majority of our users have access to state-of-the-art 8m telescopes; 
European cooperation and coordination have developed enormously; and the 
stand-alone paradigm for NOT has become obsolete. An International Evaluation 
of the NOT in 2006 supported this view 
(see report at http:/www.not.iac.es/news/reports/). 

Given NOTSA's active participation in both the OPTICON and ASTRONET European
networks (see http://www-astro.opticon.org and http://www.astronet-eu.org), 
it was natural to view the role of the NOT as part of a future pan-European 
facility of modern 2-4m telescopes. Through a series of discussions in the 
user community, this vision was turned into 
a specific long-term strategy (Andersen, 2006) and ``Development Plan'' 
(Augusteijn, 2007). 

The overarching long-term goal is to optimise the performance of the NOT 
by specialising its performance in specific areas, in concert with similar 
moves at other telescopes. This only works if an overall plan exists, and 
the ASTRONET Board is therefore appointing a European Telescope Strategy 
Review 
Committee with the charge to consider innovative ways to plan, equip, and 
operate a set of European 2-4m telescopes so as to achieve optimum scientific 
returns and cost-effectiveness, including better coordination with other 
disciplines such as space astronomy. The NOT is committed to becoming part 
of such a new, common European 2-4m facility.

Meanwhile, we continue to specialise and optimise the NOT for high-impact 
Nordic science. This includes a thorough review of the instrumentation, 
operating modes, and scheduling of projects, with emphasis on remaining
competitive in studies of transient and variable sources. As part of this 
strategy, we are seeking to move to a single set of permanently mounted 
instruments, which would enable us to respond quickly and flexibly to 
new events and optimise scientific productivity in areas where the NOT can
still be competitive if deployed intelligently. Detector and data acquisition
systems are being upgraded in parallel.

\begin{acknowledgement}
Thanks to Thomas Augusteijn for valuable comments.
\end{acknowledgement}

\bibliographystyle{}
\bibliography{}

\end{document}